# SEDNet: Shallow Encoder Decoder Network for Brain Tumor Segmentation

Chollette C. Olisah
School of Engineering
University of the West of England
Bristol, England

Sofie Van Cauter
Ziekenhuis Oost Limburg,
University Hospitals Leuven,.
Limburg, Belgium

## Abstract

*Despite the advancement in computational modeling towards brain tumor segmentation, of which several models have been developed, it is evident from the computational complexity of existing models that performance and efficiency under clinical application scenarios are still limited. Therefore, this paper proposes a tumor segmentation framework. It includes a novel shallow encoder and decoder network named SEDNet for brain tumor segmentation. The highlights of SEDNet include sufficiency in hierarchical convolutional downsampling and selective skip mechanism for cost-efficient and effective brain tumor semantic segmentation, among other features. The preprocessor and optimization function approaches are devised to minimize the uncertainty in feature learning impacted by nontumor slices or empty masks with corresponding brain slices and address class imbalances as well as boundary irregularities of tumors, respectively. Through experiments, SEDNet achieved impressive dice and Hausdorff scores of 0.9308 %, 0.9451 %, and 0.9026 %, and 0.7040 mm, 1.2866 mm, and 0.7762 mm for the non-enhancing tumor core (NTC), peritumoral edema (ED), and enhancing tumor (ET), respectively. This is one of the few works to report segmentation performance on NTC. Furthermore, through transfer learning with initialized SEDNet pre-trained weights, termed SEDNetX, a performance increase is observed. The dice and Hausdorff scores recorded are 0.9336%, 0.9478%, 0.9061%, 0.6983 mm, 1.2691 mm, and 0.7711 mm for NTC, ED, and ET, respectively. With about 1.3 million parameters and impressive performance in comparison to the state-of-the-art, SEDNet(X) is shown to be computationally efficient for real-time clinical diagnosis. The code is available on Github[1].*

## 1. Introduction

The brain tumor is an irregular growth in the brain. Whether benign or malignant, tumors can potentially pose a risk to surrounding brain tissues [1]. A benign tumor exhibits a slow development, whereas a malignant tumor exhibits abrupt development with some degrees of aggressiveness [2]. Glioblastoma is the most common and fatal type of these tumors and constitutes approximately 49.1% of all diagnosed brain tumors [3] with a median survival rate between 12 to 15 months [4]. Therefore, identifying, isolating, and quantifying subregions of the brain considered a tumor is important for early diagnosis of cancer and ensuring effective treatment planning.

During the preoperative assessment of brain tumors via magnetic resonance imaging (MRI), which is the most widely used imaging technique in neuro-oncology [5], neuroradiologists diagnose based on the structural characteristics, enhancement patterns, and the amount of surrounding edema [6]. Despite the informative details MRI reveals, the interpretation and resulting diagnosis are still dependent on the experiences and sometimes assumptions of neuroradiologists which are highly prone to cognitive bias [7]. A discrepancy in neuroradiologist interpretations accounts for approximately 60–80% of missed intracranial neoplastic abnormalities and 20–40% of misinterpretations of intracranial neoplastic abnormalities [8]. To bridge this diagnostic gap in clinical diagnosis, research in recent decades has been geared towards the computational diagnosis of brain tumors to aid accurate and timely diagnosis.

Tumors vary in size, appearance, shape, and location [9], [10] and have irregular, unclear, or discontinuous boundaries[2] due to their heterogeneous nature and imaging artifacts. Computational models can characterize complex systems due to their underlying mathematical principles and are better at minimizing any cognitive bias in diagnosis. This capability makes computational modelling a useful technique neurologists can apply in real time for effective tumor monitoring, assessment, and quantification [11]. However, the performance, reliability, and efficiency of computational models in real-time clinical application scenarios are dependent on several factors: 1) computational complexity, 2) validity and usefulness in clinical diagnosis, and 3) ability to answer clinical and radiological challenges, as mentioned by Dr. Sofie Van

---

[1] https://github.com/chollette/SEDNet_Shallow-Encoder-Decoder-Network-for-Brain-Tumor-Segmentation.git

[2] https://healthcare-in-europe.com/en/news/challenges-in-brain-tumour-segmentation.html

Cauter, a practicing neuroradiologist. In the bid to address any of these impending factors on the applicability of computational models in real-world clinical settings for the diagnosis of brain tumors, efforts have been made by several researchers toward computational model design and development [12]-[16]. Models for locating and predicting the grades of tumors [12] in order to minimize the time constraints associated with biopsy and manual inspection. Others are, models for understanding patterns and characteristics of tumor progression [13], [14], predicting survival [15], and/or estimating the size of a tumor [16]. These are all essential tasks of tumor analysis for providing patients with effective therapeutic management and treatment strategies. However, most of these tasks are reliant on the outcome of tumor segmentation. This necessitates the emphasis of this paper on the task of segmentation.

Convolutional neural networks (CNNs) are increasingly advancing brain tumor segmentation research due to their ability to efficiently discover patterns in images and their strong generalizability to similar patterns in unseen images. The BraTS dataset is a complex volumetric data series (2012–2023) with numerous slices per patient and multiple image planes (axial, coronal, and sagittal). Researchers have approached the segmentation of brain tumors with the BraTS data series from several perspectives. 3D U-Net with ResNet and Inception [17] – [19], 3D U-Net with attention mechanism [20] - [30], 2D U-Net variants [31] - [34]. The 2D U-Net variants, which by convention of its dimension should be less computationally intensive still have large trainable parameters, and more so are the transformer variants [35] - [40]. The recent advances of U-Net with attention mechanisms and transformers have reduced Hausdorff scores tremendously compared to other U-Net variants. Yet, their dice scores have not significantly improved, and the computational complexity of these models, of which some are over 30 M, is far from real-time clinical usefulness. Even though a recent work scaled the transformer network to 4.47 M trainable parameters by adopting a single encoder with a lightweight shifted multi-layer perceptron [41], it sacrificed performance for efficiency. Overall, it is evident that the performance and efficiency of existing models are limited due to their computational cost, which invariably affects their validity and usefulness in clinical diagnosis. This shortcoming takes us back to the drawing board to analyse the backbone of these models, the U-Net. U-Net was originally designed for a multi-contextual segmentation task and brain tumors are mostly localised in the brain. Therefore, tumor segmentation might not benefit greatly from the extended depths of the U-Net structure.

The localized nature of brain tumors is a significant cue for designing a brain tumor segmentation architecture which has so far not been considered. We theorise that, the more localised an object, the shallower the network structure. Therefore, this paper proposes a shallow encoder and decoder network named SEDNet for brain tumor segmentation. The proposed network is designed to consist of sufficient hierarchical convolutional blocks in the encoding pathway capable of learning the intrinsic features of brain tumors in brain slices and a decoding pathway with a selective skip path sufficient for capturing miniature local-level spatial features alongside the global-level features of brain tumors. In addition to SEDNet, several other significant contributions are made to address the high variance in appearance and shape of tumors, coupled with the ambiguity of tumor boundaries. They are as follows:

- A robust 2D shallow encoder and decoder network (SEDNet). It uses sufficient hierarchical convolutional blocks for the encoding pathway and selective skip paths for the decoding pathway to address computational complexity while simultaneously achieving impressive tumor segmentation accuracy.
- A substantially effective preprocessing algorithm for minimizing signal randomness or the proportion of noise to the signal of feature maps introduced by MRI slices containing images of a brain with a corresponding empty mask.
- Priority weighted binary cross entropy soft Dice loss (WBCESDp) is proposed as an effective learning optimization approach to address the effect of class imbalance and tumor boundary irregularity while improving learning and minimizing premature model convergence.
- Initialize a SEDNet pre-trained weight, termed SEDNetX, for the BraTS data series and show that the data volume does not limit transfer learning when the data is highly specific to a given task with minimal randomness and is transferred for the same task.

The rest of this paper is organized as follows. the proposed system is presented in Section 2. Section 3 demonstrates the experimental settings, experiments, results, and discussions, while the conclusion is provided in Section 4.

## 2. Methodology

### 2.1. Proposed system

The proposed system is designed to include a preprocessor to enable more tumor signals than noise to be captured, a robust segmentation architecture, SEDNet, to minimize the computational cost of tumor segmentation while simultaneously increasing segmentation performance, and a robust optimization function to minimize the effect of class imbalance and tumor boundary/irregularity on performance.

```
Algorithm 1: Training Preprocessor
INPUT:  FLAIR slices per sample //original 155 slices per sample
OUTPUT: non-empty, tumor slices per sample
-----------------------------Phase I----------------------------------
//Remove empty and nontumor slices
k ← (open, close)  //kernels for open and close
r ← [ ]
for f in sample do
   for i in image list do
      i = resize image i
      m = morphologyopen(i, k)
      m = morphologyclose(m, k, repeat)
      x ← find maximum of m
      if x == 0 then
         r ← create list //add files to list
      else
         c ← find Contours
         a = compute area of c
         if a ≤ T then //T, threshold for non-tumor area
            r ← create list / add more files to list
         end if
      end if
   end for
end for
//Remove files
for j in r do
   remove j
end for
-----------------------------Phase II---------------------------------
//find the number of slices per sample
L ← [ ]
for f in sample do
   total ← get length of f
   L ← set num
end for
least num = find minimum of L //find the minimum
-----------------------------Phase III--------------------------------
//let number of slices be set to the least slices of all samples
images = [ ]
masks  = [ ]
count ← 0
for f sample do
   file ← set slices path
   count ← increment by 1
   if count > least num do
      image = get name from image list
      mask = get name from mask list
      remove image
      remove mask
   end if
end for
```

## 2.2. Preprocessing

The MRI slices mostly contain images of a brain with a corresponding empty mask, which can introduce some randomness or increase the proportion of noise to the signal of feature maps during training. Therefore, to ensure that the proposed network focuses on learning brain slices with corresponding tumor signals, a preprocessing pipeline is proposed. When the network learns the salient features that best describe the tumor structure, shape, and low-level information, it will be capable of generating feature maps of rich tumor representations for discriminating between tumor and nontumor features. The proposed preprocessing pipeline consists of three phases, which are presented in Algorithm 1 (training data). The training preprocessor is focused on identifying and discarding slices of a brain with no corresponding tumor mask in phase 1. The second and third phases are optional. The choice can be to consider tumor slices either on a patient-by-patient basis or as a single entity in comparison to all other tumors. If the former is of interest, then the second and third phases are relevant; otherwise, both can be negligible. Another preprocessing step that applies to both the training and testing sets is the normalization of the data to the range [0, 1]. The ground truth labels for each brain slice are converted to binary labels to retrieve the NTC, ED, and ET labels and are further combined as a three-channel image.

## 2.3. SEDNet and Objective Function

SEDNet is designed for segmenting tumor regions of the brain. The architecture is schematically shown in Fig. 1. Although the proposed network is inspired by U-Net, it takes into consideration the localized nature of brain tumors which does not necessitate very deep networks. The proposed network is designed to consist of a hierarchical encoding pathway that can learn the intrinsic features of brain tumors in brain slices and a decoding pathway with a selective skip path sufficient for capturing miniature local-level spatial features alongside the global-level features of brain tumors. The goal of SEDNet is to reduce the number of parameters and computations for validity and usefulness in real-time while simultaneously increasing tumor segmentation performance.

*2.3.1  Encoding Pathway*

The SEDNet takes as input an MRI slice, $X \in \mathbb{R}^{W \times H \times C}$ where W is the width, H is the height and C is the channel. The encoder extracts a hierarchical feature representation of brain tumors using three downsampling convolutional blocks with $3 \times 3$ filter dimensions, reserved image dimensions, and a nonlinear activation function, ReLU, followed by $3 \times 3$ maxpooling layers. Each convolution block consists of two convolutional layers that reduce the feature maps as learning progresses with increasing filter depth starting at 32 to better learn the local-level and global-level spatial features of brain tumors. Then, a bottleneck convolutional block is added. It is set to twice $W \times H$ of the input map, X The resulting dimension of the feature map of the encoding pathway is $\frac{W}{4}, \frac{H}{4}, 2^8 C$.

*2.3.2  Decoding pathway*

On the argument that the brain tumor task presents a problem whereby the single-level contextual information is localized in the input map, a selective skip path is designed. The decoding pathway comprises two upsampling convolution blocks: an upsampling layer of $2 \times 2$ filter dimensions, a convolutional layer of $2 \times 2$ filter dimensions which replaces the U-Net transposed convolution, reserved image dimension, nonlinear activation ReLU followed by a skip path. The selected skip paths are feature maps of the first two convolutional blocks concatenated to the upsampled feature maps of deeper layers to preserve fine scale local-level and global-level spatial features of the brain tumor, respectively. These blocks better preserve tumor features, as shown in Fig. 2. Each upsampling block is preceded by a convolution block, each of which comprises two convolutional layers of the same filter dimension and depth with reserved image dimension, and nonlinear activation ReLU. The resulting dimension of the feature map of the decoding pathway is $W, H, 2^5 C$. The output layer is a convolution of $1 \times 1$ filter dimension with a depth equal to the number of pixel-wise classes, and a sigmoid activation function to output a prediction, $Y \in \mathbb{R}^{W \times H \times C}$ where $C$ is a three-channel corresponding to NTC, ED, and ET.

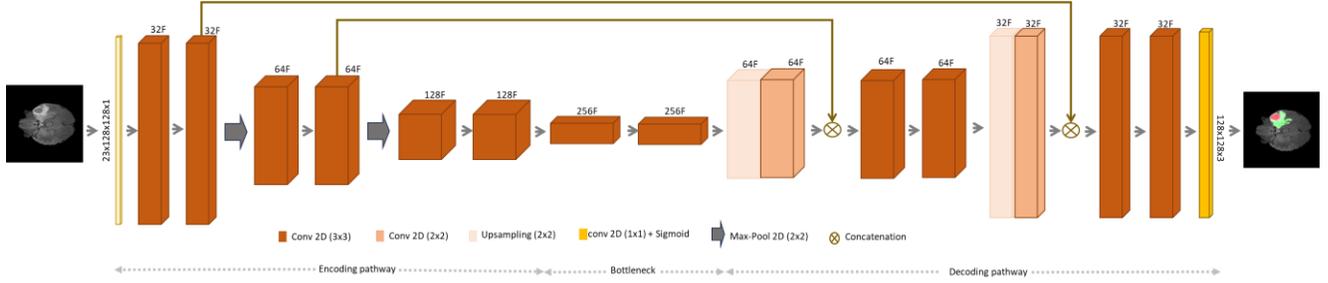

Fig. 1. The proposed architecture of 2D SEDNet.

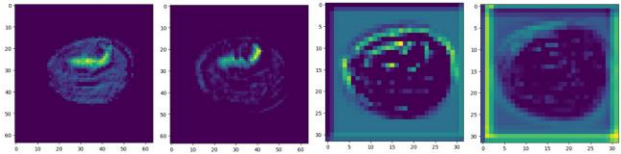

Fig. 2. Information from convolution blocks of SEDNet encoding pathway. The samples represent the output of the first, second, third, and bottleneck convolution blocks which are filter depths of 32, 64, 128, and 256.

#### 2.3.3 Transfer Learning with SEDNet

Unlike the popular opinion of transfer learning in the literature, which uses models trained with substantial amounts of data (in millions) and of distinct class labels such as pre-trained weights, this paper theorizes that the data volume does not limit transfer learning when the data are highly specific to a given task with minimal randomness and transferred for the same task. Therefore, this paper further utilizes the saved weights of SEDNet from initial learning, pre-trained, on tumor samples for transfer learning. The learned weights are initialized as a backbone for capturing the semantics of tumor regions. Then, the layers of the SEDNet architecture are frozen, while the classification layer is used as a feature extraction and learning mechanism to yield a prediction $Y_T \in \mathbb{R}^{W \times H \times C}$, where $Y_T$ is the output of transfer learning. With this approach, the SEDNet becomes SEDNetX. Transfer learning is highly important for tumor segmentation because it expands the possibility of application across the BraTS series since they share similar features. Thus, SEDNetX serves as a pre-trained weight for any of the BraTS series datasets.

#### 2.3.4 Objective Function

Since tumors take on irregular, unclear, and discontinuous boundaries due to their heterogeneous nature coupled with imaging artifacts due to different imaging sources, it is therefore appropriate to choose a loss function that can achieve a good performance. However, with consideration of the high-class imbalance of tumors of the BraTS series which can affect a good loss function as well as inhibit the performance of a well-achieving deep learning model [42], considerations of the type of loss function are necessary. Therefore, an experiment is needed to determine the most appropriate combination of BCE, $L$, and SD, $S$. The combinations considered are the binary cross entropy loss combined with soft dice loss (BCESD), equally weighted binary cross entropy loss combined with soft dice loss (WBCESD$^e$), priority weighted binary cross entropy loss combined with soft dice loss (WBCESD$^p$). The BCE is chosen to alleviate the class imbalance problem and the SD is preferred for computing the overlap between predicted ground truth labels with a bit of softness to accommodate near boundary labels resulting from tumor irregularity and discontinuity. The BSC and SD are mathematically expressed in Eq. 1 and Eq. 2, respectively.

$$L(Y_T, g) = \frac{1}{N} \sum_{i,j}^N \left[ g_{i,j} \left( \log \left( Y_{T_{i,j}} \right) \right) + (1 - g_{i,j}) \cdot \left( \log \left( 1 - Y_{T_{i,j}} \right) \right) \right] \quad (1)$$

$$S(Y_T, g) = -\frac{1}{N} \sum_{i,j}^N \left( \frac{2 \times \sum_{i,j}^N Y_{T_{i,j}} * g_{i,j} + \epsilon}{\sum_{i,j}^N \left( Y_{T_{i,j}} * g_{i,j} \right) + \sum_{i,j}^N (Y_{T_{i,j}} * g_{i,j}) + \epsilon} \right) \quad (2)$$

where $p_{i,j}$ is the predicted label and $g_{i,j}$ is the ground truth label at the $i^{th}, j^{th}$ point of $p$ and $g$ and if a perfect match the points will correspond to a 1, otherwise, it is a 0. Usually, $\epsilon$ is added to overcome division by zero error during computation.

The objective function for the proposed SEDNet architecture is their combination (Eq. 1 and Eq. 2), which is expressed as:

$$O(Y_T, g) = L(Y_T, g) + S(Y_T, g) \quad (3)$$

$$O^e(Y_T, g) = w_a^e * L(Y_T, g) + w_b^e S(Y_T, g) \quad (4)$$

$$O^p(Y_T, g) = w_a^p * L(Y_T, g) + w_b^p S(Y_T, g) \quad (5)$$

where $w_{a,b}^e$ and $w_{a,b}^p$ are equal and prioritization weights. With $w_{a,b}^p$, more weight is given to $L$ than $S$ on the assumption that the class imbalance problem grossly affects the optimization function, $O^{(e,p)}$, more than tumor boundary irregularity does. Therefore, it is expected to perform better than the other combination of BCE and SD.

## 3. Experimental settings and results

### 3.1. Experimental settings

*3.1.1  Data*

The dataset is a part of the MICCAI conference brain tumor segmentation challenge, BraTS2020. The dataset contains four tumor modalities, T1, T1-gad, T2, and FLAIR, which were manually segmented to generate the ground truth label corresponding to the whole tumor, (WT), NTC, ED, and ET, using the same labeling protocol [43]. However, only the NTC, ED, and ET are the relevant segmentation classes in this paper. The slices are preprocessed by co-registering them to the same anatomical template and interpolating them to 1mm$^3$ isotropic resolution, and then skull stripping them. The subsets of the data publicly available are the training and validation datasets, which contain 369 and 125 cases, respectively, though no ground truth label is available for the validation set.

*3.1.2  Experimental parameters*

The FLAIR modality encompasses all the tumor classes which means it is sufficient for learning the features of a tumor. Consequently, the noise and the computational cost of using all four modalities can be minimized. By applying the preprocessing algorithm on the axial plane of the 3D MRI BraTS2020 dataset, the size of each slice changes from $155 \times 240 \times 240 \times 4$ to $23 \times 128 \times 128 \times 1$, where 155 and 23 are the number of slices, 240 and 128 are the width and height of each slice, while 4 and 1 represent the modalities. This paper only utilized a single modality, the flair. The input to the network is a batch of 23 slices for each sample. Since only the training set contains ground truth labels, it was split into three – training, validation, and testing sets in the ratio 80:10:10. Then, the validation set was used for the purpose of testing for generalization. For training, the proposed SEDNet, uses the proposed loss function with ADAM optimizer and an initial learning rate of 0.0003. The adaptive learning rate was adopted and configured to gradually decrease the learning rate with a shrinkage factor of 0.3 when no improvement in the validation loss is recorded within 2 epochs. The SEDNet was trained for 50 epochs for generating weights fed into SEDNetX, retrained for 30 epochs under the same settings.

*3.1.3  Evaluation Metric*

Given the ground truth label and predicted label, the validity of the proposed model pixel-wise classification of a tumor based on their classes can be measured using the dice similarity coefficient (DSC), $Z$, [42] and the symmetric Hausdorff distance (HD) [44], $H$, are computed as:

$$Z(Y_T, g) = \frac{2 \times \sum_{i,j}^{N} |Y_{T_{i,j}} * g_{i,j}| + \epsilon}{\sum_{i,j}^{N} (Y_{T_{i,j}}^2 + g_{i,j}^2) + \epsilon} \quad (6)$$

$$H(Y_T, g) = max\{\breve{H}(Y_T, g), \breve{H}(g, Y_T)\} \quad (7)$$

where $\breve{H}$ is given as $\{min_{n \in g}\{\|m, n\|\}\}$, and where $\|.,.\|$ is the Euclidean distance function.

Since brain tumor takes on varying sizes, appearance, shape, [9,10], with irregular, unclear and discontinuous boundaries it becomes necessary to exploit metrics capable of handling these uncertainties, and literature shows that the DSC and HD are examples of such metrics. The DSC measures the overlap between, $p$ and $g$, pixels belonging to the tumor and inclusive of those pixels at the boundary [42], while HD is the maximum degree of mismatch between them [44]. A higher DSC score and lower HD translate to a perfect match.

*3.1.4  Implementation Resource*

The Tensorflow-based deep learning library, Keras, was used in a Microsoft OS environment hosting NVIDIA's Geforce RTX 3070 (16 GB RAM), Tensorflow-gpu 2.10.0, CUDA v11.6, and cuDNN v8.4.0.

### 3.2. Results and discussion

From the experiments, it is evident that even with minimal parameters and computational cost, high pixel-wise classification performances are achieved with SEDNet as shown in Fig. 3, Fig. 4, Fig. 5, Fig. 6, and Table 1 and Table 2.

An interesting observation can be seen in Fig. 3. It shows SEDNet training and validation loss curves for NTC, ED, and ET, respectively. Most loss curves of brain tumor pixel-wise classification in the literature show massive oscillations. This might signify that there exists some randomness or noisy signals in the feature maps, or that the skip connections from the deeper levels of the encoding pathway were introducing noise to the upsampled signals.

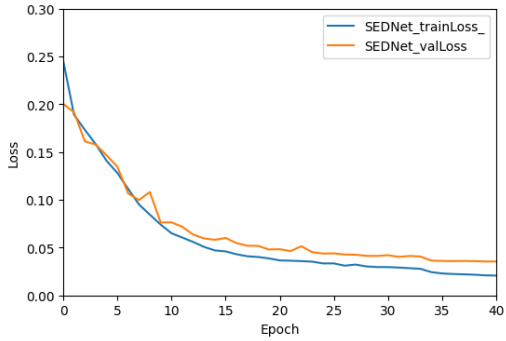

Fig 3: SEDNet training loss per epoch.

Table 1: Performance analysis of combinations of BCE and SD loss functions in the proposed framework

| Loss Function | Dice | | |
| --- | --- | --- | --- |
| | NTC | ED | ET |
| BCE | 0.9143 | 0.9376 | 0.8592 |
| BCED | 0.9279 | 0.9432 | 0.9001 |
| BCESD | **0.9315** | 0.9432 | 0.9014 |
| WBCESD$^e$ | 0.9272 | **0.9451** | 0.8875 |
| WBCESD$^p$ | 0.9308 | **0.9451** | **0.9026** |

The SEDNet, on the contrary, showed smooth curves from the first epoch to the 40th epoch. As can be observed, an increase in epoch decreases the training and validation loss and increases the dice scores. Therefore, it is relevant to state that the proposed system 1) sufficiently minimized the proportion of noise to signal of feature maps through the preprocessing pipeline and 2) the selective skip path was sufficient for capturing miniature local-level spatial features alongside the global-level features necessary for achieving tumor segmentation success at reduced parameters and computation cost.

To appreciate the contribution of the proposed combination of the BCE and SD, as the optimization function in the proposed system, several experiments were carried out and reported in Table 1. It included experiments with BCESD, WBCESD$^e$, WBCESD$^p$, and other loss functions such as the binary cross entropy loss (BCE) and binary cross entropy loss combined with dice loss (BCED). The experiment was necessary for determining the combination of BCE and SD that best optimizes learning and converges to a global minimum. As can be observed, WBCESD$^p$ impacted so much on SEDNet's ability to address class imbalance and tumor boundary irregularities. The WBCESD$^p$ loss function was observed to have better convergence performance without massive oscillations. In comparison to other loss functions experimented, WBCESD$^p$ achieved a 0.9308 dice score for NCT, which is an increase of 0.165%, 0.29%, and 0.36%, for BCE, BCED, WBCESD$^e$, respectively, but a decrease of 0.07% for BCESD. Then, a dice score of 0.9451 which is an increase of 0.75%, 0.19%, 0.19%, and 0.00% for BCE, BCED, BCESD, and WBCESD$^e$, individually. For ET, a score of 0.9026 can be observed which is an increase of 4.34%, 0.25%, 0.12%, and 1.51% for BCE, BCED, BCESD, and WBCESD$^e$, respectively. Generally, these results confirm that SEDNet achieves impressive performance irrespective of the loss function. However, the WBCESD$^p$ stood out and therefore was chosen as the loss function for the proposed segmentation system. Therefore, based on the experimental results of WBCESD$^p$, it can be theorized that class imbalance indeed grossly affects the optimization function, $O^{(e,p)}$, more than tumor boundary irregularities do. This is based on WBCESD$^p$ performance compared to the other combinations.

The impressive performance of SEDNet is further elaborated in Table 2 where the dice scores and Hausdorff scores are presented. As can be observed, the training and validation dice coefficient scores for NTC, ED, and ET are closely related and surprisingly performed equivalently when evaluated on the unseen test set. Unlike the Hausdorff distance score recorded for NTC, ED, and ET in literature which are mostly in the interval (3,9), that is, $\{H | 3 < H < 9\}$, SEDNet achieved minimal Hausdorff distances in the interval (0.5,1.29). The intriguing part of this score is that it extends to the unseen percentage of the BraTS2020 reserved for testing the generalizability of the model on similar samples. Interestingly, the SEDNet dice score also supersedes existing models in literature scores which are usually on average in the interval (0.7,0.8), that is, $\{Z | 0.7 < Z < 0.8\}$. It achieved dice scores in the interval (0.90,0.97) expressed as: $\{Z | 0.90 < Z < 0.97\}$. These results further show that the proposed system, which encompasses an efficient preprocessing pipeline and a robust SEDNet architecture impacted by the selective skip path, is significant to brain tumor segmentation.

On the BraTS2020 set designated for testing, SEDNetX can be observed in Table 2 to advance the dice score performance of SEDNet by 0.28%, 0.27%, 0.35% for NTC, ED, and ET, respectively. The interval is in the range (0.90,0.97) which takes the form, $\{Z | 0.90 < Z < 0.97\}$. Further analysis of SEDNetX's dice scores is achieved using box plot and scatter plot as shown in Fig. 4. The substantial gain in performance observed with SEDNetX shows that transfer learning can be achieved in a contrary fashion than it has long been understood and presented in the literature. In essence, SEDNetX clearly evidences that data volume in millions and of distinct class labels are not more important than highly specific data of a given task with minimal randomness, and transferred for the same task. SEDNetX achieved Hausdorff distance scores in the interval (0.5,1.28), which is expressed as $\{H | 0.5 < H < 1.28\}$ across training, validation, and testing scores recorded for NTC, ED, and ET.

Since SEDNetX improved the performance of SEDNet,

Table 2: Performance comparison of state-of-the-art and SEDNet(X).

| Model | Data set | Dice | | | Hausdorff | | | Training Parameters (M) |
|---|---|---|---|---|---|---|---|---|
| | | NTC | ED | ET | NTC | ED | ET | |
| SEDNet | Training | 0.9641 | 0.9721 | 0.9493 | 0.5842 | 0.9952 | 0.6219 | |
| | Validation | 0.9287 | 0.9454 | 0.9004 | 0.7184 | 1.2967 | 0.7786 | **1.38** |
| | Testing | 0.9308 | 0.9451 | 0.9026 | 0.7040 | 1.2866 | 0.7762 | |
| SEDNetX | Training | 0.9701 | 0.9780 | 0.9588 | 0.5455 | 0.9087 | 0.5747 | |
| | Validation | 0.9314 | 0.9483 | 0.9048 | 0.7122 | 1.2800 | 0.7716 | |
| | Testing | 0.9336 | **0.9478** | **0.9061** | **0.6983** | **1.2691** | **0.7711** | |
| dResUNet [17] | Testing | 0.8357 | - | 0.8004 | - | - | - | 30.47 |
| DPAFNet [30] | | - | 0.7810 | 0.8320 | - | - | - | - |
| Lightweight2DUnet+Inception [31] | | **0.9460** | - | 0.8860 | 9.240 | - | 3.4640 | 7.50 |
| ETUNet [40] | | - | 0.8100 | 0.8520 | - | 7.8950 | 6.0070 | 16.26 |
| SwinTransformer [41] | | - | 0.7909 | 0.6211 | - | - | - | 4.47 |

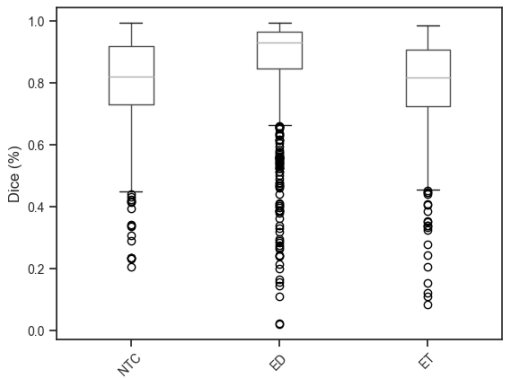
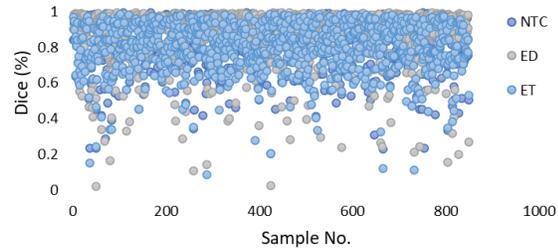

Fig. 4  Visualisation of SEDNet(X) dice score performance using boxplot (a) and scatter plot (b).

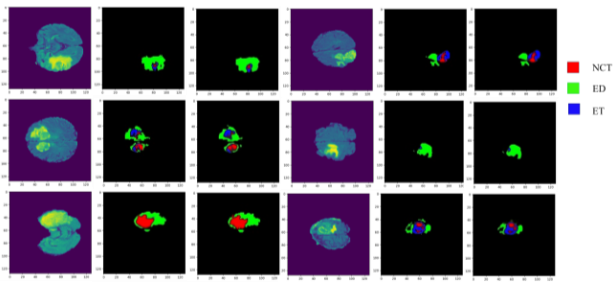

(a)

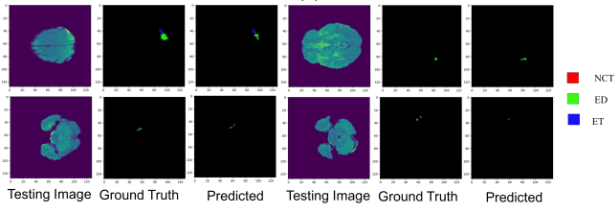

(b)

Fig. 5. Visual evaluation of SEDNet(X). (a) close match examples and (b) far-match for small-resolution tumors.

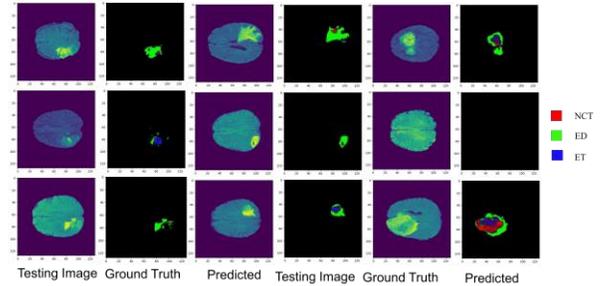

Fig. 6. Visual assessment of SEDNet(X) prediction performance on validation set.

it became the basis for visual evaluation. Fig. 5 (a) and (b) depict the visual results of SEDNetX in comparison to the ground truth (GT) labels. Precisely, Fig. 5 (a), shows that the predicted tumors are a close match to the GT tumors. On the contrary, small-resolution tumors, that is, miniaturized tumors, challenged the performance of SEDNetX as can be seen in Fig. 5 (b). However, some of these examples can be categorized as nontumors and were possibly missed by the preprocessing algorithm. On the validation set, as shown in Fig. 6, the segmentation also results appears to closely match the tumor regions as shown in the corresponding MRI slices.

## 4. Conclusion

This paper proposed a tumor segmentation system that comprises a shallow encoder-decoder network named SEDNet, a preprocessor, and an optimization function. SEDNet architecture design included sufficient hierarchical convolution blocks in the encoding and decoding pathways with selective skip paths. Through the transfer learning with initialized SEDNet pre-trained weights, SEDNetX was born. SEDNet(X) on the BraTS2020 dataset achieved dice and Hausdorff scores of 0.9336, 0.9478, 0.9061, and 0.6983, 1.2691, 0.7711 for NTC, ED, and ET, respectively. With about 1.3 million parameters and impressive performance in comparison to the state-of-the-art, SEDNet(X) is computationally efficient for real-time clinical diagnosis. While (1) and (2) of the impending factors that necessitate application of computational models to tumor analysis have been addressed in this paper, (3) is still outstanding, and therefore future work will include efforts to utilize the model to answer clinical and neuroradiological challenges. Additionally, since most of the instances where SEDNet(X) results vary from the GT were likely not tumors, a neuroradiologist analysis for defining non-tumor slices is necessary.